\documentclass[11pt,a4paper]{scrartcl}
\pdfoutput=1
\usepackage{jheppub}

\newcommand{\hhref}[1]{\href{http://arxiv.org/abs/#1}{arXiv:#1}}

\title{Testing Localization in Neutrino Oscillations}

\author{Dmitry~V.~Zhuridov}

\affiliation{Scuola Normale Superiore, and INFN, Piazza dei Cavalieri 7, 56126 Pisa, Italy}

\emailAdd{dmitry.zhuridov@sns.it, jouridov@mail.ru}

\abstract{The neutrino wave packet localization in short-baseline neutrino oscillation experiments, such as MiniBooNE, is investigated. 
It is shown that the transition from localization to delocalization may be observed for large neutrino mass splitting of order 1\,eV, 
e.g., in theories with sterile neutrinos.}


\arxivnumber{1203.2764}

\begin{document}

\maketitle

\section{Introduction}

The neutrino oscillations were theoretically introduced more than half a century ago~\cite{Pontecorvo,MNS} and experimentally confirmed more than ten years ago~\cite{PDG2010}. 
However the plane wave approach to the neutrino oscillations
leads to some ``paradoxical'' issues, such as the relevance of the ``same energy'' and ``same momentum'' assumptions, etc. All these theoretical issues 
can be resolved~\cite{paradoxes_nu_osc} in more generic wave packet approach, which was developed in Refs.~\cite{Nussinov,Kayser1981,Giunti-Kim-Lee,Rich} 
and many other works. 
Two important necessary conditions for the neutrino oscillations are the neutrino coherence and localization. The former means that the distance traveled by neutrinos 
should be smaller than the distance over which the wave packets corresponding to different mass eigenstates separate due to the difference of their group velocities. 
The latter means that the neutrino wave packet should be localized better than the oscillation length. 

In this paper we consider the observability of the transition from neutrino localization to delocalization with the corresponding averaging out of the neutrino 
oscillation in short-baseline experiments, such as LSND~\cite{LSND} and 
MiniBooNE~\cite{MiniBooNE1,MiniBooNE,MiniBooNE2011}. We show that it is possible to observe this transition simply by changing the decay pipe length 
in the case of large neutrino mass squared difference $\Delta m^2\sim1\,\text{eV}^2$ 
and mixing $\sin^22\theta\sim0.1$, e.g., in the models with sterile neutrinos.
We discuss the neutrino wave packets and their sizes in the next section and investigate neutrino localization in short-baseline experiments 
in section~\ref{section:localization}.

\section{Neutrino wave packets}\label{section:general}

Consider a flavor neutrino eigenstate $\nu_\alpha$, which was born at the moment $t=0$ in the point ${\bf x}=0$ of a source.
The wave packet describing the evolved state at a point $(t,{\bf x})$ is~\cite{paradoxes_nu_osc}
\begin{eqnarray}
 |\nu_\alpha(t,{\bf x})\rangle = \sum_iU_{\alpha i}^*\Psi_i(t,{\bf x})|\nu_i\rangle,
\end{eqnarray}
where $U$ is the lepton miximg matrix, and the wave packet of a free propagating neutrino $\nu_i$ of definite mass $m_i$, which has momentum ${\bf p}_i$, is
\begin{eqnarray}
 \Psi_i(t,{\bf x}) = \int \frac{d^3p}{(2\pi)^{3/2}} f_i^S({\bf p}-{\bf p}_i) e^{i{\bf px}-iE_i(p)t},
\end{eqnarray}
where $E_i(p)=\sqrt{p^2+m_i^2}$ with $p\equiv|{\bf p}|$, and $f_i^S({\bf p}-{\bf p}_i)$ is the momentum distribution 
function, which is sharply peaked at or very close to ${\bf p}={\bf p}_i$, with the width of the peak (uncertainty of the neutrino momentum) 
$\sigma_p\ll p_i$ and the superscript ``$S$'' corresponding to the source.

The spatial size $\sigma_x$ of the wave packet of
neutrino $\nu_i$ created in decay of a particle with the width $\Gamma$ and the lifetime $\tau$ is~\cite{1201.4128}
\begin{eqnarray}
 \sigma_{x} \sim \sigma_p^{-1} \sim \Gamma^{-1} \sim \tau.
\end{eqnarray} 
If the particle with relativistic $\gamma$ 
factor decays in flight, its lifetime and the corresponding size of the neutrino wave packet are dilated, so that $\sigma_x\sim\gamma\tau$.  
However if this particle decays in a decay tunnel with length $L_S$ 
than $\sigma_x \sim \text{min}(\gamma\tau,L_S)$, see section~8.3 of Ref.~\cite{Giunti-Kim}. 
For the neutrino created in a decay or scattering process occurring in a medium $\sigma_{x} \sim \text{min}(\tau_X)$,
where $\tau_X$ is the average time between two collisions in the medium for a particle $X$, participating in the neutrino creation process. 
In the following we consider the accelerator neutrino experiments, in which the medium effect is negligible, such as MiniBooNE  
with the neutrinos produced in 50 m decay pipe filled with air.

The velocity of spreading of the wave packet can be estimated as
\begin{eqnarray}
 v \sim \frac{\partial^2E_i}{\partial p_i^2}\sigma_p \sim  \frac{m_i^2}{E_i^3}\sigma_x^{-1} \sim   10^{-25}
\left(\frac{1\,\text{MeV}}{E}\right)^3 \left(\frac{m_i}{1\,\text{eV}}\right)^2  \left(\frac{1\,\text{m}}{\sigma_x}\right).
\end{eqnarray}
Hence spreading of the neutrino wave packets can be typically neglected. 

\section{Neutrino localization}\label{section:localization}

The $\nu_\alpha\to\nu_\beta$ oscillation probability at the distance $L$ from the source can be written as
\begin{eqnarray}\label{eq:osc_probability}
 P_{\nu_\alpha\to\nu_\beta}(L) = \sum_i|U_{\alpha i}|^2|U_{\beta i}|^2 + 2\sum_{j>i} \text{Re}\left[U_{\alpha j}^*U_{\alpha i}U_{\beta j}U_{\beta i}^* 
e^{-2\pi i\frac{L}{L_\text{osc}^{ij}}}\right]  
\end{eqnarray}
with the oscillation length
\begin{eqnarray}\label{eq:L_osc}
 L_\text{osc}^{ij}=\frac{4\pi p}{\Delta m_{ji}^2}.
\end{eqnarray}

Denote $\sigma_L=\min(L_\text{S},L_\text{D},\sigma_x)$ 
with the characteristic size $L_\text{S(D)}$ of the neutrino source (detector) in the direction of neutrino propagation. 
If $\sigma_L$ is relatively large, it characterizes the uncertainty of $L$ in Eq.~\eqref{eq:osc_probability}.  
Hence, for $\sigma_L\sim L_\text{osc}^{ij}$ the correspondent oscillation 
term in Eq.~\eqref{eq:osc_probability} will be averaged out, which is called delocalization of neutrinos.
Sterile neutrinos with small enough $L_\text{osc}$ (large $\Delta m^2$) may provide the opportunity to observe the 
transition from localized to delocalized neutrinos in the short-baseline accelerator neutrino experiments with large enough neutrino source and detector.

\begin{center}
 \begin{figure}[tb]
 \centering
\includegraphics[width=0.95\textwidth]{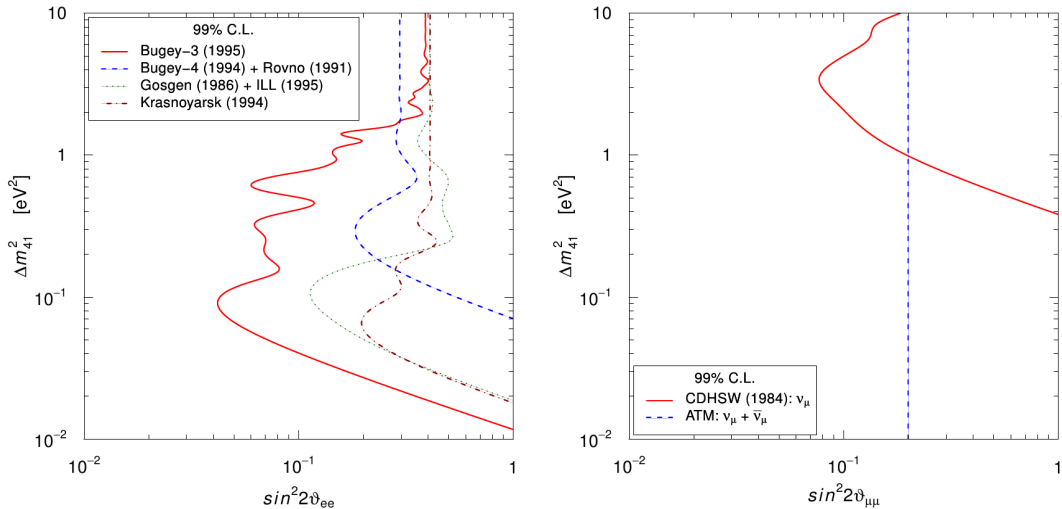}
   \caption{Exclusion curves obtained in Ref.~\cite{Giunti_sterile_nu} from the data of reactor $\bar\nu_e$ disappearance experiments~\cite{Mention:2011rk},
CDHSW $\nu_\mu$ disappearance experiment~\cite{Dydak:1983zq}, and from atmospheric neutrino data~\cite{Maltoni:2007zf}.}
   \label{fg:sterile_nu}
 \end{figure}
\end{center}

Consider the extension of usual three-neutrino mixing with the addition of one massive (sterile) neutrino. The survival 
probability in short-baseline experiments in this 3+1 neutrino mixing scheme~\cite{Okada:1996kw,Bilenky:1996rw,Bilenky:1999ny,Maltoni:2004ei} 
can be written as~\cite{Giunti_sterile_nu}
\begin{eqnarray}\label{eq:P_sterile}
 P_{\nu_\alpha\to\nu_\alpha}(L) \approx   1 -  \frac{1}{2}\sin^22\theta_{\alpha\alpha} \left[ 1 - \cos\left(\frac{\Delta m_{41}^2 L}{2E}\right)\right], 
\end{eqnarray}
where
$\sin^22\theta_{\alpha\alpha} = 4|U_{\alpha4}|^2 (1-|U_{\alpha4}|^2)$. 
The exclusion ranges for $\Delta m_{41}^2$ versus $\sin^22\theta_{ee}$ and $\sin^22\theta_{\mu\mu}$ are shown in Fig.~\ref{fg:sterile_nu}.

\begin{center}
 \begin{figure}[tb]
 \centering
\includegraphics[width=0.45\textwidth]{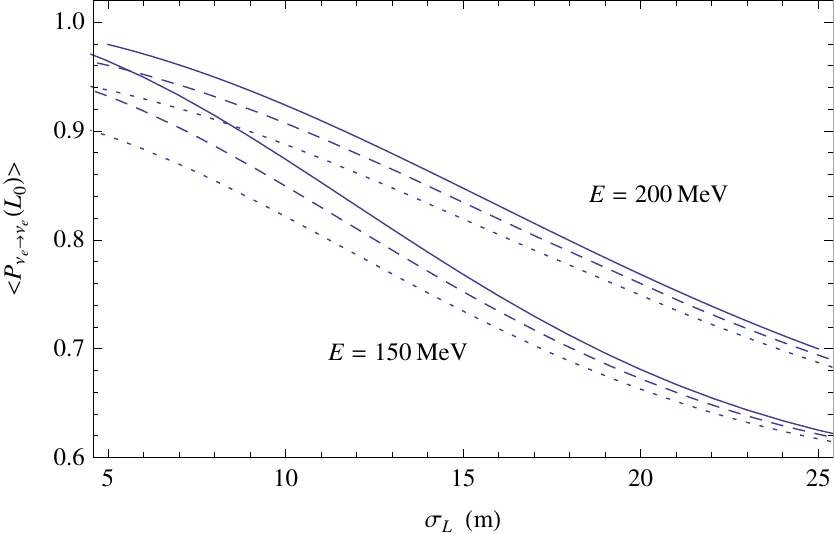}
\includegraphics[width=0.45\textwidth]{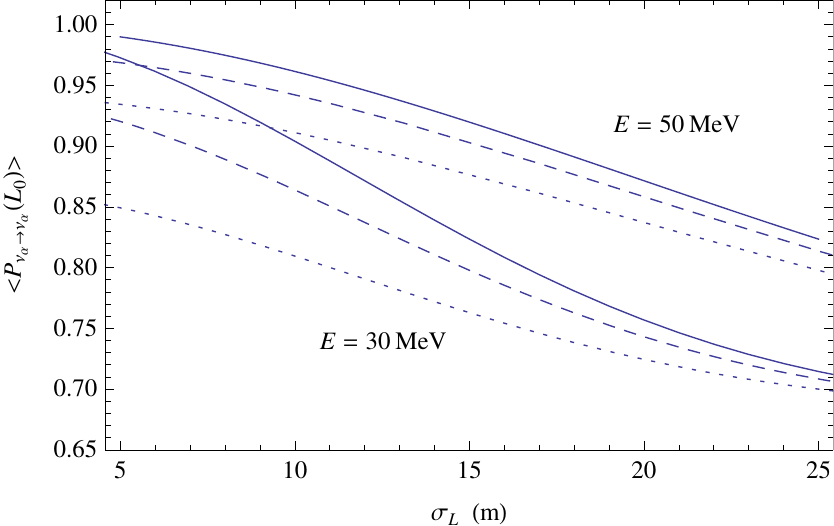}
   \caption{$\langle P_{\nu_e\to\nu_e}(L_0)\rangle$ versus  $\sigma_L$ for $\Delta m^2=5$\,eV$^2$ and 
$\sin^22\theta=0.3$ (left); and $\langle P_{\nu_\alpha\to\nu_\alpha}(L_0)\rangle$ versus  $\sigma_L$ for $\Delta m^2=1$\,eV$^2$ and  
$\sin^22\theta=0.2$ (right).}
   \label{fg:Pee_averaged}
 \end{figure}
\end{center}

Denote by $L_0=L_\text{osc}^{14}$ the coordinate of the first nontrivial maximum of the oscillation curve. (In general, one can choose one of the first 
extremums, for which the decoherence effect is small, while the effect of averaging out of the neutrino oscillation due to the delocalization can be spectacular.) 
The mean survival probability $\langle P_{\nu_\alpha\to\nu_\alpha}(L_0)\rangle$, which can be measured at $L_0$, 
can be found by the standard averaging of Eq.~\eqref{eq:P_sterile} with the Gaussian distribution (see section 7.6 of Ref.~\cite{Giunti-Kim})
\begin{eqnarray}
 \phi\left(\frac{L}{E}; L_0, \langle E\rangle\right) =  \frac{1}{\sqrt{2\pi \sigma_{L/E}^2}} \text{exp}\left[ -\frac{(L/E-L_0/\langle E\rangle)^2}{2\sigma_{L/E}^2} \right] 
\end{eqnarray}
with 
the standard deviation
\begin{eqnarray}
 \sigma_{L/E} = \frac{L_0}{\langle E \rangle} \sqrt{\left(\frac{\sigma_L}{L_0}\right)^2 + \left(\frac{\Delta E}{\langle E \rangle}\right)^2},
\end{eqnarray}
where $\Delta E$ is the energy resolution of a detector.  
For averaging the cosine in Eq.~\eqref{eq:P_sterile} we use the approximation 
\begin{eqnarray}
 \left<\cos\left(\frac{\Delta m_{41}^2 L}{2E}\right)\right>_{L_0/\langle E\rangle}  
&=&  \int\limits_0^\infty \cos\left(\frac{\Delta m_{41}^2 L}{2E}\right) 
\phi\left(\frac{L}{E}; L_0, \langle E\rangle\right) d\frac{L}{E}   \\
&\approx&   \int\limits_{-\infty}^\infty \cos\left(\frac{\Delta m_{41}^2 L}{2E}\right) 
\phi\left(\frac{L}{E}; L_0, \langle E\rangle\right) d\frac{L}{E}   \\
&=&   \cos\left(\frac{\Delta m_{41}^2 L_0}{2\langle E\rangle}\right) \text{exp}\left[-\frac{1}{2} 
\left(\frac{\Delta m_{41}^2}{2}\sigma_{L/E}\right)^2\right],
\end{eqnarray}
which is still very good for relatively large values of $\sigma_L$ in Fig.~\ref{fg:Pee_averaged}.

Consider a short-baseline neutrino experiment with the neutrinos created in the decays of free mezons or muons flying in the decay pipe, and 
large neutrino detector centered at $L_0$. Suppose $L_S\leq L_D$, so that $\sigma_L$ can be controlled at the source site simply by changing the decay pipe length. 
Fig.~\ref{fg:Pee_averaged} (left) 
shows $\langle P_{\nu_e\to\nu_e}(L_0)\rangle$ versus $\sigma_L$ calculated for $\Delta m_{41}^2=5$\,eV$^2$, 
$\sin^22\theta_{14}=0.3$, and the values of neutrino energy  $E=200$\,MeV (three upper lines) 
and 150\,MeV (three lower lines), for which $L_0=99.2$ and 74.4 m, respectively. 
The solid, dashed and dotted lines represent $\Delta E=0$, 10 and 15 MeV, respectively. 
Fig.~\ref{fg:Pee_averaged} (right) 
shows $\langle P_{\nu_\alpha\to\nu_\alpha}(L_0)\rangle$ ($\alpha=e,\mu$) versus $\sigma_L$ for $\Delta m_{41}^2=1$\,eV$^2$, 
$\sin^22\theta_{14}=0.2$, and the neutrino energy  $E=50$ and 30\,MeV, for which $L_0=124.0$ and 74.4 m, respectively. 
The solid, dashed and dotted lines represent $\Delta E=0$, 3 and 5 MeV, respectively. 
Fig.~\ref{fg:Pee_averaged} demonstrates the averaging out of the first maximum of neutrino oscillation with increasing of the decay pipe length.

We note that neutrinos with masses of few eV may explain the dark matter in halos around galaxies~\cite{Nicolaidis:2012qs}.

In conclusion, the possibility to test the transition from neutrino localization to delocalization in the short-baseline accelerator 
neutrino experiments is demonstrated.

\section*{{Acknowledgements}}

The author thanks Anatoly Borisov (Lomonosov MSU) for useful discussions and comments. 
This work was supported in part by the EU ITN ``Unification in the LHC Era'',
contract PITN-GA-2009-237920 (UNILHC) and by MIUR under contract 2006022501.


\end{document}